**Addressing Extreme Propensity Scores in Estimating Counterfactual Survival Functions via the Overlap Weights**[1]

**Authors:** Chao Cheng[1,2], MS; Fan Li[3,4,5], PhD; Laine E. Thomas[4,5], PhD; and Fan (Frank) Li[1,2], PhD

**Correspondence to** Fan (Frank) Li, PhD, Department of Biostatistics, Yale School of Public Health, Suite 200, Room 229, 135 College Street, New Haven, Connecticut 06510 (email: fan.f.li@yale.edu)

**Author affiliations:**
[1]Department of Biostatistics, Yale School of Public Health, New Haven, Connecticut
[2]Center for Methods in Implementation and Prevention Science, Yale School of Public Health, New Haven, Connecticut
[3]Department of Statistical Science, Duke University, Durham, North Carolina
[4]Duke Clinical Research Institute, Durham, North Carolina
[5]Department of Biostatistics and Bioinformatics, Duke University School of Medicine, Durham, North Carolina

**Running head:** Overlap Weighting with Survival Outcomes

---

**Abbreviations:**

CDF: cumulative distribution function

IPCW: inverse probability of censoring weights

IPTW: inverse probability of treatment weighting

OW: overlap weighting

PS: propensity score

**Abstract**

The inverse probability of treatment weighting (IPTW) approach is popular for evaluating causal effects in observational studies, but extreme propensity scores could bias the estimator and induce excessive variance. Recently, the overlap weighting approach has been proposed to alleviate this problem, which smoothly down-weights the subjects with extreme propensity scores. Although advantages of overlap weighting have been extensively demonstrated in literature with continuous and binary outcomes, research on its performance with time-to-event or survival outcomes is limited. In this article, we propose estimators that combine propensity score weighting and inverse probability of censoring weighting to estimate the counterfactual survival functions. These estimators are applicable to the general class of balancing weights, which includes IPTW, trimming, and overlap weighting as special cases. We conduct simulations to examine the empirical performance of these estimators with different propensity score weighting schemes in terms of bias, variance, and 95% confidence interval coverage, under various degree of covariate overlap between treatment groups and censoring rate. We demonstrate that overlap weighting consistently outperforms IPTW and associated trimming methods in bias, variance, and coverage for time-to-event outcomes, and the advantages increase as the degree of covariate overlap between the treatment groups decreases.

## INTRODUCTION

Survival or time-to-event outcomes are common in comparative effectiveness research and require unique handling because they are usually incompletely observed due to right-censoring. In observational studies, a popular approach to draw causal inference with survival outcomes is to apply the Cox proportional hazards model to the inverse probability weighted sample to estimate the adjusted survival curves [1] or the causal hazard ratio [2]. However, the Cox model assumes the hazards ratio is constant over time, which is often violated and leads to biased estimates. Moreover, when the Cox model includes covariates, the resulting coefficient of treatment variable can only be interpreted conditionally rather than marginally [3]. On the other hand, using a Cox model without covariates would fail to accommodate covariate-dependent censoring. The second approach is to combine inverse probability of treatment weighting (IPTW) with the pseudo-observation to estimate causal survival function [4]. Each pseudo-observation is treated as the uncensored contribution to the target parameter and allows for standard method to proceed as if the outcomes are completely observed. However, pseudo-observations are jackknife statistics and require intensive computation for estimating the survival functions with large sample sizes. A final method is to combine IPTW with inverse probability of censoring weight (IPCW), which is intrinsically connected to the Kaplan-Meier estimator [5], to estimate the counterfactual survival functions while accommodate covariate-dependent censoring.

As a key component of the aforementioned methods, IPTW weights each subject proportional to the inverse of its probability of receiving the observed treatment—the propensity score (PS)—given pre-treatment covariates [6-8] and estimates the causal effect by differencing the weighted average of the outcomes between treatment groups. However, IPTW is unstable when there is a lack of overlap in the covariates between groups, reflected by the presence of PS values close to 0 or 1 [9-12]. These PS tails lead to extreme inverse probability weights, resulting in bias and excess variance in the causal estimates [9]. The two main methods that mitigate the numerical instability of IPTW are PS trimming [9, 10, 13] and overlap weighting (OW) [12, 14, 15]; both focus on a subpopulation with adequate covariate overlap. Crump et al. [9] developed the symmetric trimming rule to reduce the variance of IPTW; Stürmer et al. [10, 16] studied the asymmetric trimming rule to address unmeasured confounding in the tails of PS distribution. Trimming methods may be sensitive to the choice of cutoff point. In contrast, OW weights each subject by its probability of being assigned to the opposite group and thus smoothly down-weights subjects as their propensity scores move toward 0 and 1, avoiding arbitrarily choosing a threshold to remove subjects.

The problem posed by extreme PSs in IPTW also applies to survival outcomes [2], and intuitively, OW would mitigate this problem. However, investigation on the empirical performance of OW with survival outcomes has been limited with two exceptions. Mao et al. [11] demonstrated that combining OW with an Cox outcome model leads to notable efficiency gain from IPTW for a range of causal estimands. Zeng et al. [4] combined OW with pseudo-observations and showed that OW leads to optimal efficiency, but their approach requires jackknifing and therefore computationally intensive. In this paper, we develop computationally efficient OW method to estimate the causal effect on a right-censored survival outcome without making assumptions about the outcome model. To simultaneously address extreme PS values and selection bias due to censoring, we combine OW and IPCW to estimate the counterfactual survival functions, and derive closed-form variance estimators. Extensive simulations are conducted to compare OW and IPTW, with and without trimming, in terms of bias, variance, and coverage.

**METHODS**

*Notations and Estimands*

Consider an observational comparative effective study with $n$ subjects and survival outcomes. For each subject $i(= 1, \dots, n)$, we observe pretreatment covariates $X_i$, and treatment status $A_i(= a)$, with $a = 1$ for treatment and $a = 0$ for control, and censoring time $C_i$. Each unit has a failure time $T_i$, which is subject to right-censoring at time $C_i$. Therefore, we only observe the survival time, $U_i = \min(T_i, C_i)$, and the censoring indicator, $\delta_i = I(T_i \leq C_i)$.

To define causal effects, we introduce notation of counterfactuals and the concept of target population. Specifically, $T_i^{(a)}$ is the potential failure time and $C_i^{(a)}$ is the potential censoring time that would be observed under treatment status $a$. In the PS balancing weight framework [6], each unit is assigned a weight that is a function of the PS, and the weighted covariate distributions in the treatment and control groups are balanced. The weighted population is referred to as the target population, on which the treatment effect is defined. Each weighting scheme $w$ within the balancing weight family (e.g. IPTW, OW, trimming) corresponds to a specific target population. Then on the target population corresponding to $w$, we can define the counterfactual survival distribution on as $S_w^{(a)}(t) = P_w(T^{(a)} > t)$ for $a = 1, 0$ (See Web Appendix 1 for technical definitions), and related average causal effect [11]:

$$\Delta_w(t) = S_w^{(1)}(t) - S_w^{(0)}(t), \quad 0 \leq t \leq t_{max} \qquad (1)$$

where $t_{max}$ is the maximum follow-up time.

This article focuses on developing unbiased estimators for $\Delta_w(t)$. Throughout, we maintain four standard assumptions in causal survival analysis [17]: (A1) consistency and no-interference [17], which ensures $T_i = T_i^{(a)}$ and $C_i = C_i^{(a)}$ for $A_i = a$; (A2) conditional exchangeability, which assumes away unmeasured confounders; (A3) covariate-dependent censoring, which assumes censoring time is independent of failure time given observed covariates in each group; and (A4) positivity, such that the conditional probability of treatment assignment is bounded away from 0 and 1, and the conditional survival probability of censoring is larger than 0.

*Propensity Score Balancing Weights*

PS is the probability of receiving treatment given covariates: $e(X) = P(A = 1|X)$. Balancing the true PS between treatment groups will balance the distribution of each covariate in expectation [18]. However, because PS is unknown in observational studies, it must be estimated, most commonly through a logistic regression $e(X, \boldsymbol{\beta}) = 1/(1 + \exp(-\boldsymbol{\beta}^T X))$. We denote estimated PS as $\hat{e}(X) = e(X, \widehat{\boldsymbol{\beta}})$. Below we describe three common PS balancing weight schemes.

*Inverse probability of treatment weighting.* The balancing weight $w_i$ given by IPTW is $1/\hat{e}(X_i)$ for treated subjects and $1/(1 - \hat{e}(X_i))$ for control subjects. The target population is the population represented by the study sample and the causal estimand, $\Delta_{IPTW}(t)$, measures the difference in the counterfactual survival probabilities over time on this overall population.

*IPTW with trimming.* There are two popular PS trimming strategies. The first is symmetric trimming [19], which excludes subjects whose estimated PS is outside the range $(\alpha, 1 - \alpha)$, with common choices of the threshold $\alpha$ being 0.1 or 0.05 [9]. The second is asymmetric trimming [10], which involves two steps. It first ensures common support by excluding subjects outside of the common PS region; then it excludes all subjects whose PS is either below the $q$ quantile of the treated subjects or above the $1 - q$ quantile of the control subjects [20].

*Overlap weighting.* OW weights each subject by the estimated probability of being assigned to the opposite treatment group [14, 21], i.e., $w_i = 1 - \hat{e}(X_i)$ for treated subjects and $w_i = \hat{e}(X_i)$ for control subjects. OW is maximized when PS is 0.5 and diminishes when PS moves towards 0 or 1. OW emphases the treatment comparison for the subpopulation of subjects at clinical equipoise, namely who have substantial probability to be assigned to either treatment groups – the usual target population of a randomized trial [12]. The OW estimand $\Delta_{OW}(t)$ targets at the treatment comparison with the optimal internal validity. OW has several advantages, including exact balance, minimum variance, and avoiding arbitrary choice of cutoff point. More details can be found in [12, 21].

### Combining PS Weighting and Inverse Probability of Censoring Weighting

To extend PS weighting to survival outcomes, one must accommodate the censoring process. Specifically, we use IPCW to address selection bias associated with the right-censoring [5]. The censoring process can be characterized by the survival distribution of the potential censoring time conditional on $X$, $K_c^{(a)}(t, X) = P(C \geq t | X, A = a)$ for $a = 1, 0$. Similar to the PS, $K_c^{(a)}(t, X)$ is considered as the censoring score, which is unknown but can be estimated from the

observed data. We estimate $K_c^{(a)}(t, X)$ in each treatment group separately to obviate the need for specifying full treatment-by-covariate interactions. While there are many methods in the survival analysis literature for estimating $K_c^{(a)}(t, X)$, including the Cox model [22] and additive risk model [23], we consider parametric survival regressions [24] for simplicity. We use the Weibull regression to model the group-specific censoring scores, $K_c^{(a)}(t, X) = \exp(-t^{\gamma_a} \exp(\boldsymbol{\theta}_a^T X))$, where $\gamma_a$ is a scale parameter and $\boldsymbol{\theta}_a$ is the coefficients associated with $X$. Accordingly, $\widehat{K}_c^{(a)}(t, X)$ is obtained by plugging in the maximum likelihood estimators, $\hat{\gamma}_a$ and $\widehat{\boldsymbol{\theta}}_a$.

We propose two unbiased estimators to combine PS weighting and IPCW for estimating the counterfactual survival functions. The first estimator (estimator I) is motivated by Satten and Datta [25], who showed that the Kaplan-Meier estimator of the survival probability can be represented as one minus an IPCW estimator of the cumulative distribution function (CDF). Building on this, we first estimate the counterfactual CDF, $F_w^{(a)}(t) = P_w(T^{(a)} \leq t)$, and then obtain $S_w^{(a)}(t) = 1 - F_w^{(a)}(t)$. Specifically, estimator I is given by

$$\widehat{\Delta}_w^I(t) = \widehat{S}_w^{(1),I} - \widehat{S}_w^{(0),I}$$

$$= \left(1 - \frac{\sum_{i=1}^n w_i A_i \delta_i I(U_i \leq t) / \widehat{K}_c^{(1)}(U_i, X_i)}{\sum_{i=1}^n w_i A_i}\right) - \left(1 - \frac{\sum_{i=1}^n w_i (1 - A_i) \delta_i I(U_i \leq t) / \widehat{K}_c^{(0)}(U_i, X_i)}{\sum_{i=1}^n w_i (1 - A_i)}\right), (2)$$

where $I(.)$ denotes an indicator function, $w_i$ is a specific balancing weight, and $\delta_i / \widehat{K}_c^{(a)}(U_i, X_i)$ is the inverse probability of censoring weights applied only among the non-censored observations through study follow-up. When the censoring process is independent of covariates, i.e., $\widehat{K}_c^{(a)}(U_i, X_i) = \widehat{K}_c^{(a)}(U_i)$, estimator I simplifies to a PS weighted Kaplan-Meier estimator [5, 25, 26]. Under covariate-dependent censoring, estimator I further allows us to flexibly incorporate covariates to address selection bias.

The second estimator (estimator II) is similar to the construction in Bai et al. [27] and directly estimates the counterfactual survival functions without first computing the CDF. Specifically, estimator II is

$$\widehat{\Delta}_w^{II}(t) = \hat{S}_w^{(1),II} - \hat{S}_w^{(0),II}$$

$$= \frac{\sum_{i=1}^n w_i A_i I(U_i > t)/\widehat{K}_c^{(1)}(t, \boldsymbol{X}_i)}{\sum_{i=1}^n w_i A_i} - \frac{\sum_{i=1}^n w_i (1-A_i) I(U_i > t)/\widehat{K}_c^{(0)}(t, \boldsymbol{X}_i)}{\sum_{i=1}^n w_i (1-A_i)}, \quad (3)$$

in which the inverse probability of censoring weights are applied to all observations regardless of their censoring status. To estimate the causal effect at time $t$, one subtle difference between estimators I and II lies in computing the censoring scores: $\widehat{\Delta}_w^I(t)$ requires the censoring score $\widehat{K}_c^{(a)}(U_i, \boldsymbol{X}_i)$ evaluated at the *observed survival time* $U_i$, and only among the non-censored units through the entire study follow-up, whereas $\widehat{\Delta}_w^{II}(t)$ requires the censoring score $\widehat{K}_c^{(a)}(t, \boldsymbol{X}_i)$ evaluated at *the time of interest t* but among all units. This subtle difference in IPCW may have implications on statistical efficiency of the resulting estimator, which we investigate in the simulations.

In Web Appendix 2, we prove that under Assumptions (A1)-(A4), both estimator I and II are pointwise consistent to $\Delta_w(t)$ under any PS balancing weight scheme $w$. In Web Appendix 3 we further show that under homoscedasticity conditions, OW achieves the smallest asymptotic pointwise variance for estimating $\Delta_w(t)$ among the class of balancing weights, for both estimator I and II. However, as we demonstrate in simulations, the homoscedasticity condition is not required for OW to exhibit improved efficiency as compared to the IPTW and PS trimming.

Several previous work with non-censored outcomes also empirically verified that the OW improves efficiency from IPTW regardless of the homoscedasticity condition [4, 14].

While the variance of $\hat{\Delta}_w^I(t)$ and $\hat{\Delta}_w^{II}(t)$ can be estimated by bootstrapping, it may be computationally intensive with large data sets. We develop closed-form variance estimators for the proposed estimators based on the M-estimation theory [28], when the PS is estimated by a logistic regression and the censoring score is by the Weibull regression. Details are given in Web Appendix 4 and 5. For symmetric and asymmetric PS trimming, we recommend refitting the PS and censoring models post trimming, and then apply our closed-form IPTW variance estimator.

**SIMULATION STUDIES**

*Simulation design*

We conduct simulation studies under a range of scenarios to compare IPTW, PS trimming, and OW, under various degree of covariate overlap between groups. We consider a sample size of $n = 2,000$, with 6 covariates, $X_1$-$X_6$. Specifically, $X_1$, $X_2$, and $X_3$ are drawn from a multivariate normal distribution, with mean zero, unit marginal variance, and pairwise correlation coefficient 0.5. We then independently simulate $X_4$, $X_5$, and $X_6$ from a Bernoulli distribution with probability of 0.5. The true PS follows a logistic model:

$$logit(e(\boldsymbol{X})) = \beta_0 + \beta_1 X_1 + \beta_2 X_2 + \beta_3 X_3 + \beta_4 X_4 + \beta_5 X_5 + \beta_6 X_6,$$

where $(\beta_1, \beta_2, \beta_3, \beta_4, \beta_5, \beta_6)^T = (0.2\psi, 0.3\psi, 0.3\psi, -0.2\psi, -0.3\psi, -0.2\psi)^T$. We choose $\psi \in (1, 3, 5)$ to represent increasing separation for the treatment-specific PS distribution, leading to strong, moderate, and weak overlap. The intercept in the PS model, $\beta_0$, was chosen such that half of the subjects receive treatment. The PS distributions under the 3 levels of covariate overlap are visualized in Web Figure 4. The counterfactual survival outcomes $T^{(a)}$ are drawn from the following Weibull distribution:

$$P(T^{(a)} > t|X) = \exp(-t^\gamma \exp(m_a(X))),$$

where the scale parameter $\gamma$ is set to 0.95 and $m_a(X)$ is a linear function of covariates $X$. We consider $m_a(X) = \alpha_0 + \alpha_1 X_1 + \alpha_2 X_2 + \alpha_3 X_3 + \alpha_4 X_4 + \alpha_5 X_5 + \alpha_6 X_6$ with $(\alpha_0, \alpha_1, \alpha_2, \alpha_3, \alpha_4, \alpha_5, \alpha_6)^T = (-1 - \eta, 0.4 - \eta, 0.2 - \eta, 0.1 - \eta, -0.1 - \eta, -0.2 - \eta, -0.3 - \eta)^T$, where we set $\eta = 0$ for the treated units ($a = 1$) and $\eta = 0.4$ for the controlled units ($a = 0$) such that treatment will benefit survival experience. The true potential survival functions $S_w^{(1)}(t)$ and $S_w^{(0)}(t)$ on the overall and overlap populations are illustrated in Web Figure 5, for each level of covariate overlap. Under strong overlap, the OW survival functions are nearly identical with the IPTW survival functions; as the covariates are increasingly separated, the OW survival functions start diverging from the IPTW survival functions. Finally, assuming no causal effect of treatment on censoring (i.e., $C^{(1)} = C^{(0)} = C$), we simulate the censoring time from an exponential distribution:

$$P(C \geq t|X) = \exp(-t \exp(\lambda - 0.3X_1 + 0.5X_2 + 0.5X_3 + 0.2X_4 - 0.4X_5 - 0.5X_6)),$$

where $\lambda$ is set to -3 and -1.6 corresponding to censoring rate at 25% and 50%, respectively. Finally, given the survival and censoring times, the observed failure time is given by $U = \min(T, C)$ with a censoring indicator $\delta = I(T \leq C)$.

We simulated 2000 datasets for each scenario. For each dataset, we focus on $\Delta_w(t)$ at four pre-specified time points $t \in (t_1, t_2, t_3, t_4)$, and use the IPTW, OW, and symmetric and asymmetric PS trimming approaches combined with either $\widehat{\Delta}_w^I(t)$ or $\widehat{\Delta}_w^{II}(t)$. Following the literature on PS trimming [10, 12], we consider $\alpha = 0.05, 0.10,$ and $0.15$ for symmetric trimming and $q = $ 0, 0.01, 0.05 for asymmetric trimming. The four time points, $(t_1, t_2, t_3, t_4)$, are specified as the time such that approximately 80%, 60%, 40%, and 20% of the participants remain at risk. Table 1 summarizes the four time points and the corresponding $\Delta_{IPTW}(t)$ and $\Delta_{OW}(t)$ at each level of covariate overlap and censoring rate. In each scenario, the true values of $\Delta_w(t)$ given by the symmetric and asymmetric trimming approaches lie between $\Delta_{IPTW}(t)$ and $\Delta_{OW}(t)$, and are not shown here. We compare the performance of these estimators in terms of three operating characteristics: relative bias, relative efficiency, and 95% confidence interval coverage. We define the percent bias as the ratio of bias and the true estimand under each weighting scheme. The relative efficiency is defined as the ratio of the Monte Carlo variance of $\widehat{\Delta}^I_{IPTW}$ to the variance for the weighted estimator in consideration. Larger relative efficiency indicates a more efficient estimator. The reproducible R code for simulations is available at:

https://github.com/chaochengstat/OW_Survival/tree/main/Simulation_Code.

*Simulation results*

Table 2 demonstrates the percent bias for each estimator with increasingly strong tails in the PS distributions under a 25% censoring rate. All the estimators have little bias under strong or moderate overlap ($\psi = 1$ and 3), but the IPTW estimators, $\widehat{\Delta}^I_{IPTW}(t)$ and $\widehat{\Delta}^{II}_{IPTW}(t)$, exhibit noticeable bias under weak overlap ($\psi = 5$). For all three choices of $\alpha$, symmetric trimming reduces the bias compared with IPTW without trimming, with all percent biases controlled within 1%. The asymmetric trimming approach with $q = 0$ also reduces bias, but sometimes its bias is sizeable with $q > 0$. In contrast, OW generally minimizes the percent bias across the levels of overlap. Within the same weighting scheme ($w$), $\widehat{\Delta}^I_w$ and $\widehat{\Delta}^{II}_w$ leads to similar estimates.

Table 3 presents the relative efficiency with a 25% censoring rate. From Table 3, OW estimators maximize the efficiency as compared to IPTW and the trimming approaches under all three levels of covariates overlap, where its efficiency advantages are more pronounced (e.g., 10 times more efficient than IPTW) when the degree of overlap decreases and at an earlier time point with a relatively smaller amount of censoring. The trimming approaches are generally less efficient than OW, but still improves efficiency compared with IPTW when the degree of overlap decreases. Between the two estimators $\widehat{\Delta}_w^I(t)$ and $\widehat{\Delta}_w^{II}(t)$, we observe that $\widehat{\Delta}_w^I(t)$ can sometimes be slightly more efficient, but the relative efficiency of $\widehat{\Delta}_w^I(t)$ compared to $\widehat{\Delta}_w^{II}(t)$ mostly fluctuates between 1 and 1.2 under all balancing weight scheme $w$ considered.

In Table 4, we present the 95% confidence interval coverage with a 25% censoring rate. Both OW and symmetric trimming, with confidence intervals calculated based on our closed-form sandwich variance estimator, provide nominal coverages under all of three degree of overlap, with both estimator I and II. IPTW and asymmetric trimming provide desirable coverages with strong or moderate overlap ($\psi = 1$ and 3), but the coverages are slightly attenuated under weak overlap ($\psi = 5$).

Finally, we examine the estimators under a more challenging scenario with a 50% censoring rate. The results are presented in Web Tables 1-3, with respect to percent bias, efficiency, and coverage, respectively. Overall, the result patterns with a 50% censoring rate are analogous to those with a 25% censoring rate. For both estimators I and II, OW provides minimal bias with nominal coverage and optimal efficiency under all three levels of overlap; symmetric trimming also provides little bias and desirable coverage with all three trimming thresholds but is less efficient than OW; the IPTW and asymmetric trimming approach still exhibit moderate biases under weak overlap. Notably, as compared to results with a 25% censoring rate, the efficiency advantage of estimator I becomes slightly more pronounced.

**EMPIRICAL STUDY WITH THE COMPARE-UF FIBROID REGISTRY**

We analyzed the data from the COMPARE-UF Fibroid Registry to compare the effectiveness between hysterectomy and myomectomy for the treatment of symptomatic leiomyomas [29, 30]. The starting population included women receiving these procedures who were over 30, not trying to get pregnant, and not-missing short-term follow-up. We further restricted to the patients receiving minimally invasive hysterectomy (506 patients) and minimally invasive myomectomy (213 patients) to avoid multiple versions of each treatment. We considered the time from procedure to return to work as a time-to-event (survival) outcome, and administratively censored it at 60 days allowing a 2-week window beyond the planned short-term follow-up at 45 days. Approximately, 10% outcomes are censored due to loss of follow up or maximum follow-up after 60 days. The PS was estimated using logistic regression adjusting for baseline characteristics including age, race, ethnicity, insurance type, time since first diagnosis of leiomyomas, bleeding symptoms, prior procedures, prior pregnancies, depression/anxiety, and UFS-QOL components scores. Unweighted baseline characteristics and weighted baseline characteristics for the combined and overlap target populations are shown in Web Tables 4 and 5, suggesting that hysterectomy patients are generally older, more likely to be multiparous, and having severer symptoms compared to myomectomy patients. Of note, such strong differences between the two patient groups can make the IPTW results for the combined population scientifically less interpretable because there were younger patients with close to zero probability to receive hysterectomy, and vice versa. The histogram of the estimated PS (Web Figure 6) also indicated moderate lack of covariate overlap between treatment groups. We compared the causal effect of minimally invasive hysterectomy versus minimally invasive myomectomy using IPTW and OW, coupled with IPCW to mitigate potential selection bias due to censoring. The censoring scores were estimated from Weibull regression within each group adjusting for the same set of covariates as the PS model.

We first report the results given by estimator I. Table 5 presents the difference in counterfactual probability to return to work under different weighting estimators in the first seven weeks (time since procedure). The 95% confidence intervals for all estimates were obtained by our closed-form variance estimators. The results consistently suggested that minimally invasive myomectomy lead to a significantly shorter time to return to work. However, a noticeable difference between IPTW and OW is that the latter is associated with a much narrower confidence interval at each time point. Panels (a) and (b) in Figure 1 present the estimated difference in counterfactual probability in returning to work with IPTW and OW, respectively, which further confirms the increased efficiency of OW by emphasizing the subpopulation at clinical equipoise. Furthermore, the confidence band of IPTW straddles zero until day 14, whereas that of OW excludes zero since day 8. The corresponding counterfactual survival functions are presented in Panels (c) and (d) in Figure 1, with similar point estimates under IPTW and OW, but much narrower confidence bands under OW. In this empirical analysis, the estimated survival functions turn out to be similar to those returned by PS weighted Kaplan-Meier curves (results omitted for brevity), implying that the censoring is likely non-informative.

We repeated the analyses using estimator II (see Web Table 6 and Web Figure 7), and found almost identical results under estimators I and II for the same weighting scheme; this is expected from our simulations due to a low censoring rate.

**DISCUSSION**

We investigate the PS weighting method for observational studies with a right-censored survival outcome. We proposed two weighting estimators that combine PS weighting and IPCW to estimate the counterfactual survival functions, which can accommodate both independent and covariate-dependent censoring. These estimators are applicable to the general class of balancing weights, including IPTW, trimming, OW, and matching weights [31]. Using extensive simulations, we compared these estimators with different weighting schemes under various degrees of covariate overlap and censoring rates. Our simulations suggest that OW is the most efficient in estimating counterfactual survival probabilities, consistently outperforming IPTW and trimming in terms of bias, variance, and coverage, and produces valid estimates with minimal bias and nominal coverage even when the covariates are highly imbalanced with a substantial censoring proportion. The advantage increases as the overlap between groups decreases. These findings demonstrate that the optimal variance property of OW among balancing weights extends to survival outcomes, a result of which we also provide theoretical justifications in Web Appendix 3.

Because a key complication of survival outcome is right-censoring, we also investigated two different approaches to combine PS weighting and IPCW. We proposed two types of estimators (estimators I and II) and proved that both are point-wise consistent and OW achieves optimal variance for both estimators under the respective homoscedasticity conditions. Our simulations indicate that the two estimators have similar operating characteristics, with estimator I being slightly more efficient than estimator II, particularly under a higher censoring rate. A possible explanation underlying this efficiency result is that estimator I directly generalizes the Kaplan-Meier estimator [25], which is the nonparametric maximum likelihood estimator for estimating survival functions in the absence of covariates. Our simulation results clearly indicate that differences in constructing IPCW can have implications for estimation efficiency, under both IPTW and OW. We generally recommend coupling estimator I with OW to address lack of overlap with survival outcomes.

We emphasize that different weighting schemes correspond to different target populations. Specifically, IPTW targets at the overall population that is represented by the study sample, whereas OW and trimming approaches target at subpopulation with sufficient overlap between treatments, that is, the subpopulation close to clinical equipoise. When the study sample has strong overlap, the corresponding true estimands of IPTW and OW have negligible differences, but they diverge as the covariates in the two groups become increasingly separated. Choosing an appropriate target population (and equivalently weighting scheme) is crucial for analyzing and interpreting the findings from an observational study. For example, when there is a severe lack of overlap between treatment groups, the investigators should question whether the overall population (which corresponds to IPTW) is the proper target. If choosing to use IPTW with trimming, one already implicitly changes the target population to the subpopulation with sufficient overlap. However, trimming requires arbitrary choosing the cutoff; in contrast, OW can be viewed as a statistically-justified continuous version of trimming, which focuses on the subpopulation whose treatment decision is the most uncertain, and for whom the causal evidence is arguably the most needed.

A potential limitation of the proposed weighting estimators is that the resulting counterfactual survival functions are not guaranteed to be monotonically non-decreasing in $t$ in finite samples. A simple modification to ensure monotonicity without compromising the asymptotic properties of the weighting estimator is to define $\hat{S}_w^*(t) = \min_{s \leq t} \hat{S}_w(s)$, where $\hat{S}_w(t)$ is the original survival functions given by estimator I or II [23]. An empirical evaluation of this strategy is beyond the scope of this article and left for future work. Finally, to facilitate application, we provide a step-by-step R tutorial in Web Appendix 6.

**ACKNOWLEDGEMENT**

This research is supported in part by the Patient-Centered Outcomes Research Institute (PCORI) contract ME-2018C2-13289. The COMPARE-UF study, which constitutes our motivating example, was supported by the Agency for Healthcare Research and Quality grant RFA-HS-14-006. The contents of this article are solely the responsibility of the authors and do not necessarily represent the view of PCORI nor AHRQ. We appreciate the clinical input and motivating questions from COMPARE-UF PI Evan Myers and COMPARE-UF investigators. The authors also thank Siyun Yang for help with the discussion and analysis of the COMPARE-UF study.

**WEB MATERIAL**

The web appendices, tables, and figures are available at

https://github.com/chaochengstat/OW_Survival/blob/main/OW_Survival_Web_Material.pdf

# REFERENCES

1. Cole, S.R. and M.A. Hernán, *Adjusted survival curves with inverse probability weights.* Computer methods and programs in biomedicine, 2004. **75**(1): p. 45-49.

2. Austin, P.C. and T. Schuster, *The performance of different propensity score methods for estimating absolute effects of treatments on survival outcomes: a simulation study.* Statistical methods in medical research, 2016. **25**(5): p. 2214-2237.

3. Hernán, M.A., *The hazards of hazard ratios.* Epidemiology (Cambridge, Mass.), 2010. **21**(1): p. 13.

4. Zeng, S., F. Li, and L. Hu. Propensity Score Weighting Analysis of Survival Outcomes Using Pseudo-observations [preprint]. *arXiv preprint*. 2021. (arXiv:2103.00605).

5. Robins, J.M. and D.M. Finkelstein, *Correcting for noncompliance and dependent censoring in an AIDS clinical trial with inverse probability of censoring weighted (IPCW) log-rank tests.* Biometrics, 2000. **56**(3): p. 779-788.

6. Lunceford, J.K. and M. Davidian, *Stratification and weighting via the propensity score in estimation of causal treatment effects: a comparative study.* Statistics in medicine, 2004. **23**(19): p. 2937-2960.

7. Xu, S., et al., *Use of stabilized inverse propensity scores as weights to directly estimate relative risk and its confidence intervals.* Value in Health, 2010. **13**(2): p. 273-277.

8. Austin, P.C. and E.A. Stuart, *Moving towards best practice when using inverse probability of treatment weighting (IPTW) using the propensity score to estimate causal treatment effects in observational studies.* Statistics in medicine, 2015. **34**(28): p. 3661-3679.

9. Crump, R.K., et al., *Dealing with limited overlap in estimation of average treatment effects.* Biometrika, 2009. **96**(1): p. 187-199.

10. Stürmer, T., et al., *Treatment effects in the presence of unmeasured confounding: dealing with observations in the tails of the propensity score distribution—a simulation study.* American journal of epidemiology, 2010. **172**(7): p. 843-854.

11. Mao, H., et al., *On the propensity score weighting analysis with survival outcome: Estimands, estimation, and inference.* Statistics in medicine, 2018. **37**(26): p. 3745-3763.

12. Li, F., L.E. Thomas, and F. Li, *Addressing extreme propensity scores via the overlap weights.* American journal of epidemiology, 2019. **188**(1): p. 250-257.

13. Walker, A.M., et al., *A tool for assessing the feasibility of comparative effectiveness research.* Comparative effectiveness research, 2013. **3**: p. 11-20.

14. Li, F., K.L. Morgan, and A.M. Zaslavsky, *Balancing covariates via propensity score weighting.* Journal of the American Statistical Association, 2018. **113**(521): p. 390-400.

15. Li, F., *Propensity score weighting for causal inference with multiple treatments.* Annals of Applied Statistics, 2019. **13**(4): p. 2389-2415.

16. Stürmer, T., et al., *Propensity Score Weighting and Trimming Strategies for Reducing Variance and Bias of Treatment Effect Estimates: A Simulation Study.* American Journal of Epidemiology, 2021.

17. Hernán, M.A. and J.M. Robins, *Causal inference: what if.* 2020, Boca Raton: Chapman & Hall/CRC.

18. Austin, P.C., *An introduction to propensity score methods for reducing the effects of confounding in observational studies.* Multivariate behavioral research, 2011. **46**(3): p. 399-424.

19. Lee, B.K., J. Lessler, and E.A. Stuart, *Weight trimming and propensity score weighting.* PloS one, 2011. **6**(3): p. e18174.

20. Yoshida, K., et al., *Multinomial extension of propensity score trimming methods: a simulation study.* American journal of epidemiology, 2019. **188**(3): p. 609-616.

21. Thomas, L.E., F. Li, and M.J. Pencina, *Overlap weighting: a propensity score method that mimics attributes of a randomized clinical trial.* Jama, 2020. **323**(23): p. 2417-2418.

22. Cox, D.R., *Regression models and life-tables.* Journal of the Royal Statistical Society: Series B (Methodological), 1972. **34**(2): p. 187-202.

23. Lin, D. and Z. Ying, *Semiparametric analysis of the additive risk model.* Biometrika, 1994. **81**(1): p. 61-71.

24. Klein, J.P. and M.L. Moeschberger, *Survival analysis: techniques for censored and truncated data.* New York: Springer; 2003.

25. Satten, G.A. and S. Datta, *The Kaplan–Meier estimator as an inverse-probability-of-censoring weighted average.* The American Statistician, 2001. **55**(3): p. 207-210.

26. Robins, J.M. and A. Rotnitzky. Recovery of information and adjustment for dependent censoring using surrogate markers. *AIDS epidemiology*. New York: Springer; 1992: 297-331.

27. Bai, X., A.A. Tsiatis, and S.M. O'Brien, *Doubly-Robust Estimators of Treatment-Specific Survival Distributions in Observational Studies with Stratified Sampling.* Biometrics, 2013. **69**(4): p. 830-839.

28. Tsiatis, A., *Semiparametric theory and missing data.* Now York: Springer; 2006.

29. Nicholson, W.K., et al., *Short-term health-related quality of life after hysterectomy compared with myomectomy for symptomatic leiomyomas.* Obstetrics & Gynecology, 2019. **134**(2): p. 261-269.

30. Laughlin-Tommaso, S.K., et al., *Short-term quality of life after myomectomy for uterine fibroids from the COMPARE-UF Fibroid Registry.* American journal of obstetrics and gynecology, 2020. **222**(4): p. 345. e1-345. e22.

31. Li, L. and T. Greene, *A weighting analogue to pair matching in propensity score analysis.* The international journal of biostatistics, 2013. **9**(2): p. 215-234.

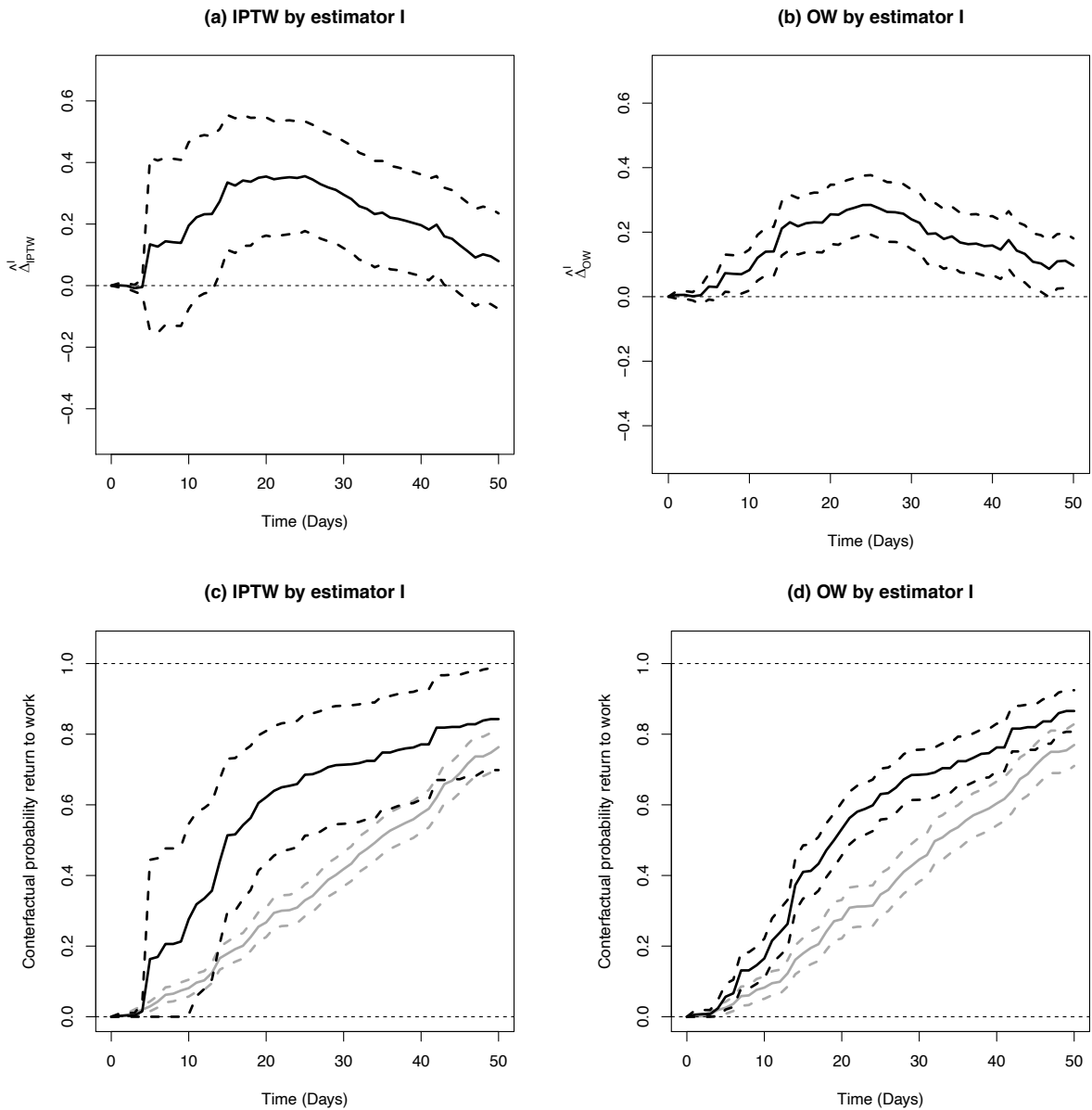

**Figure 1.** Panels (a) and (b): Estimated differences in counterfactual probability in returning to work between minimally invasive myomectomy and minimally invasive hysterectomy in the first 50 days. Panels (c) and (d): Estimated counterfactual probabilities in returning to work between minimally invasive myomectomy (black lines) and minimally invasive hysterectomy (gray lines) in the first 50 days. All the results are calculated based on estimator I. We plot the curves up to 50 days since there is no event in the minimally invasive myomectomy group after day 49. The 95% confidence intervals were obtained with the proposed closed-form variance estimators.

**Table 1**. The four time points $(t_1, t_2, t_3, t_4)$ with their corresponding $\Delta_{IPTW}$ and $\Delta_{OW}$ at strong, moderate, and weak levels of covariate overlap under 25% and 50% censoring rates[a].

| Covariates Overlap | 25% Censoring Rate | | | 50% Censoring Rate | | |
|---|---|---|---|---|---|---|
| | $\psi = 1$ | $\psi = 3$ | $\psi = 5$ | $\psi = 1$ | $\psi = 3$ | $\psi = 5$ |
| $t_1$ | 1.032 | 1.180 | 1.234 | 0.728 | 0.802 | 0.827 |
| $\Delta_{IPTW}(t_1)$ | 0.154 | 0.168 | 0.173 | 0.120 | 0.129 | 0.132 |
| $\Delta_{OW}(t_1)$ | 0.148 | 0.153 | 0.154 | 0.115 | 0.115 | 0.114 |
| $t_2$ | 2.604 | 2.919 | 3.027 | 1.768 | 1.919 | 1.969 |
| $\Delta_{IPTW}(t_2)$ | 0.255 | 0.266 | 0.269 | 0.213 | 0.222 | 0.225 |
| $\Delta_{OW}(t_2)$ | 0.251 | 0.259 | 0.263 | 0.208 | 0.209 | 0.210 |
| $t_3$ | 5.174 | 5.667 | 5.829 | 3.358 | 3.588 | 3.662 |
| $\Delta_{IPTW}(t_3)$ | 0.299 | 0.300 | 0.300 | 0.277 | 0.282 | 0.284 |
| $\Delta_{OW}(t_3)$ | 0.299 | 0.310 | 0.316 | 0.275 | 0.281 | 0.284 |
| $t_4$ | 10.383 | 11.074 | 11.227 | 6.350 | 6.658 | 6.751 |
| $\Delta_{IPTW}(t_4)$ | 0.270 | 0.262 | 0.261 | 0.300 | 0.299 | 0.298 |
| $\Delta_{OW}(t_4)$ | 0.273 | 0.282 | 0.288 | 0.301 | 0.312 | 0.318 |

[a] Values of $\Delta_{IPTW}$ and $\Delta_{OW}$ are evaluated using a sufficient large sample with 500,000 observations.

**Table 2**. Percent bias of the estimators in the presence of increasingly strong tails in the propensity score distribution (censoring rate = 25%).

| Estimator | | $\widehat{\Delta}^I_w(t)$ | | | | $\widehat{\Delta}^{II}_w(t)$ | | | |
|---|---|---|---|---|---|---|---|---|---|
| | | $t_1$ | $t_2$ | $t_3$ | $t_4$ | $t_1$ | $t_2$ | $t_3$ | $t_4$ |
| **Overlap Weighting** ($\psi = 1$) | | -0.1 | -0.2 | -0.1 | -0.2 | 0.0 | -0.2 | -0.2 | -0.1 |
| ($\psi = 3$) | | -0.4 | -0.2 | -0.2 | -0.3 | -0.4 | -0.2 | -0.3 | -0.2 |
| ($\psi = 5$) | | -0.7 | -0.2 | 0.1 | -0.2 | -0.9 | -0.4 | -0.1 | -0.2 |
| **IPTW** | | | | | | | | | |
| No Trimming | ($\psi = 1$) | -0.1 | -0.2 | -0.1 | -0.1 | 0.0 | -0.2 | -0.2 | -0.1 |
| | ($\psi = 3$) | -0.9 | -0.8 | -0.4 | -0.5 | -0.9 | -0.7 | -0.4 | -0.3 |
| | ($\psi = 5$) | -5.1 | -3.0 | -1.2 | -0.7 | -4.6 | -2.9 | -1.5 | -1.1 |
| Symmetric Trimming | | | | | | | | | |
| $\alpha = 0.05$ | ($\psi = 1$) | -0.1 | -0.2 | -0.1 | -0.1 | 0.0 | -0.2 | -0.2 | -0.1 |
| | ($\psi = 3$) | -0.5 | -0.1 | -0.2 | -0.3 | -0.6 | -0.2 | -0.3 | -0.3 |
| | ($\psi = 5$) | -1.0 | -0.4 | -0.2 | -0.4 | -1.2 | -0.5 | -0.3 | -0.2 |
| $\alpha = 0.1$ | ($\psi = 1$) | -0.1 | -0.2 | -0.1 | -0.1 | 0.0 | -0.2 | -0.2 | -0.1 |
| | ($\psi = 3$) | -0.5 | -0.2 | -0.3 | -0.2 | -0.6 | -0.2 | -0.4 | -0.2 |
| | ($\psi = 5$) | -0.8 | -0.2 | 0.1 | 0.1 | -0.8 | -0.3 | 0.0 | 0.1 |
| $\alpha = 0.15$ | ($\psi = 1$) | -0.2 | -0.2 | 0.0 | 0.0 | -0.2 | -0.3 | -0.2 | 0.0 |
| | ($\psi = 3$) | -0.4 | -0.3 | -0.4 | -0.3 | -0.5 | -0.4 | -0.5 | -0.2 |
| | ($\psi = 5$) | -0.8 | -0.2 | 0.1 | 0.0 | -0.9 | -0.2 | 0.0 | 0.0 |
| Asymmetric Trimming | | | | | | | | | |
| $q = 0$ | ($\psi = 1$) | -0.2 | -0.2 | 0.1 | 0.2 | -0.2 | -0.2 | -0.1 | 0.2 |
| | ($\psi = 3$) | 1.1 | 1.6 | 2.1 | 2.0 | 1.0 | 1.5 | 2.0 | 2.1 |
| | ($\psi = 5$) | 1.9 | 3.9 | 5.3 | 4.6 | 2.3 | 3.8 | 4.9 | 4.0 |
| $q = 0.01$ | ($\psi = 1$) | -3.4 | -0.8 | 1.3 | 2.7 | -3.4 | -0.9 | 1.1 | 2.7 |
| | ($\psi = 3$) | -6.3 | -0.9 | 3.4 | 6.2 | -6.5 | -1.0 | 3.1 | 6.2 |
| | ($\psi = 5$) | -4.7 | -0.7 | 2.4 | 4.4 | -5.0 | -0.9 | 2.4 | 4.5 |
| $q = 0.05$ | ($\psi = 1$) | -8.0 | -2.1 | 3.1 | 6.6 | -7.9 | -2.1 | 2.9 | 6.7 |
| | ($\psi = 3$) | -5.7 | -1.1 | 3.0 | 5.7 | -5.6 | -1.1 | 3.0 | 5.8 |
| | ($\psi = 5$) | -2.8 | -0.4 | 2.1 | 2.9 | -2.9 | -0.6 | 1.9 | 3.3 |

**Table 3**. Relative efficiency of the estimators relative to the Original Approach IPW estimator in the presence of increasingly strong tails in the propensity score distribution (censoring rate = 25%).

| Estimator | | $\widehat{\Delta}_w^I(t)$ | | | | $\widehat{\Delta}_w^{II}(t)$ | | | |
|---|---|---|---|---|---|---|---|---|---|
| | | $t_1$ | $t_2$ | $t_3$ | $t_4$ | $t_1$ | $t_2$ | $t_3$ | $t_4$ |
| **Overlap Weighting** ($\psi = 1$) | | 1.08 | 1.00 | 0.98 | 1.00 | 1.00 | 0.94 | 0.91 | 0.87 |
| ($\psi = 3$) | | 3.75 | 2.79 | 2.28 | 2.29 | 3.42 | 2.58 | 2.13 | 2.12 |
| ($\psi = 5$) | | 11.45 | 8.10 | 6.39 | 6.46 | 10.22 | 7.27 | 5.73 | 6.04 |
| **IPTW** | | | | | | | | | |
| No Trimming | ($\psi = 1$) | 1.00 | 1.00 | 1.00 | 1.00 | 0.94 | 0.94 | 0.94 | 0.89 |
| | ($\psi = 3$) | 1.00 | 1.00 | 1.00 | 1.00 | 0.95 | 0.99 | 1.05 | 1.2 |
| | ($\psi = 5$) | 1.00 | 1.00 | 1.00 | 1.00 | 0.99 | 0.97 | 1.07 | 1.33 |
| Symmetric Trimming | | | | | | | | | |
| $\alpha = 0.05$ | ($\psi = 1$) | 1.00 | 1.00 | 1.00 | 1.00 | 0.94 | 0.94 | 0.94 | 0.89 |
| | ($\psi = 3$) | 2.59 | 2.20 | 1.90 | 1.89 | 2.38 | 2.05 | 1.82 | 1.88 |
| | ($\psi = 5$) | 7.95 | 5.73 | 4.66 | 5.04 | 7.24 | 5.34 | 4.53 | 4.83 |
| $\alpha = 0.1$ | ($\psi = 1$) | 1.00 | 1.00 | 1.00 | 1.00 | 0.93 | 0.94 | 0.93 | 0.88 |
| | ($\psi = 3$) | 3.09 | 2.29 | 1.94 | 2.03 | 2.85 | 2.12 | 1.81 | 1.83 |
| | ($\psi = 5$) | 8.78 | 6.30 | 4.98 | 5.33 | 7.99 | 5.86 | 4.62 | 4.74 |
| $\alpha = 0.15$ | ($\psi = 1$) | 1.02 | 0.98 | 0.98 | 0.99 | 0.95 | 0.93 | 0.92 | 0.87 |
| | ($\psi = 3$) | 3.01 | 2.17 | 1.82 | 1.93 | 2.72 | 2.02 | 1.70 | 1.70 |
| | ($\psi = 5$) | 8.43 | 5.88 | 4.64 | 4.87 | 7.78 | 5.50 | 4.29 | 4.31 |
| Asymmetric Trimming | | | | | | | | | |
| $q = 0$ | ($\psi = 1$) | 0.99 | 0.99 | 0.98 | 1.00 | 0.92 | 0.94 | 0.92 | 0.88 |
| | ($\psi = 3$) | 0.99 | 0.94 | 0.92 | 0.95 | 0.94 | 0.93 | 0.95 | 1.09 |
| | ($\psi = 5$) | 0.91 | 0.88 | 0.87 | 0.88 | 0.89 | 0.86 | 0.93 | 1.14 |
| $q = 0.01$ | ($\psi = 1$) | 1.01 | 0.94 | 0.94 | 0.98 | 0.94 | 0.88 | 0.87 | 0.85 |
| | ($\psi = 3$) | 2.74 | 2.12 | 1.81 | 1.92 | 2.56 | 1.97 | 1.70 | 1.79 |
| | ($\psi = 5$) | 8.34 | 5.66 | 4.66 | 5.29 | 7.50 | 5.31 | 4.39 | 4.82 |
| $q = 0.05$ | ($\psi = 1$) | 0.92 | 0.81 | 0.8 | 0.84 | 0.86 | 0.76 | 0.74 | 0.74 |
| | ($\psi = 3$) | 2.60 | 1.86 | 1.55 | 1.59 | 2.40 | 1.74 | 1.46 | 1.40 |
| | ($\psi = 5$) | 6.84 | 4.52 | 3.72 | 3.89 | 6.20 | 4.19 | 3.36 | 3.31 |

**Table 4**. Coverage rate of the 95% confidence intervals for the estimators in the presence of increasingly strong tails in the propensity score distribution (censoring rate = 25%).

| Estimator | | $\widehat{\Delta}_w^I(t)$ | | | | $\widehat{\Delta}_w^{II}(t)$ | | | |
|---|---|---|---|---|---|---|---|---|---|
| | | $t_1$ | $t_2$ | $t_3$ | $t_4$ | $t_1$ | $t_2$ | $t_3$ | $t_4$ |
| **Overlap Weighting** ($\psi = 1$) | | 94.8 | 94.9 | 94.9 | 95.4 | 95.3 | 95.4 | 95.6 | 95.2 |
| ($\psi = 3$) | | 95.1 | 95.0 | 94.3 | 95.9 | 95.5 | 95.5 | 95.1 | 96.4 |
| ($\psi = 5$) | | 94.9 | 95.5 | 94.9 | 96.1 | 95.3 | 95.3 | 95.5 | 96.1 |
| **IPTW** | | | | | | | | | |
| No Trimming | ($\psi = 1$) | 95.5 | 95.2 | 95.1 | 95.6 | 95.6 | 95.2 | 95.5 | 95.1 |
| | ($\psi = 3$) | 94.8 | 96.1 | 95.1 | 95.7 | 93.9 | 94.6 | 94.7 | 94.8 |
| | ($\psi = 5$) | 91.3 | 94.3 | 94.6 | 94.5 | 91.4 | 93.4 | 93.6 | 93.2 |
| Symmetric Trimming | | | | | | | | | |
| $\alpha = 0.05$ | ($\psi = 1$) | 95.5 | 95.2 | 95.1 | 95.6 | 95.5 | 95.2 | 95.5 | 95.2 |
| | ($\psi = 3$) | 94.3 | 96.0 | 95.3 | 95.3 | 93.9 | 95.0 | 95.3 | 95.8 |
| | ($\psi = 5$) | 94.5 | 95.3 | 94.8 | 96.5 | 95.2 | 94.9 | 95 | 95.9 |
| $\alpha = 0.1$ | ($\psi = 1$) | 95.6 | 95.2 | 95.3 | 95.6 | 95.5 | 95.3 | 95.3 | 95.3 |
| | ($\psi = 3$) | 95.7 | 95.4 | 94.6 | 96.9 | 95.1 | 94.9 | 94.8 | 96.6 |
| | ($\psi = 5$) | 95.0 | 95.9 | 95.1 | 96.9 | 95.4 | 96.0 | 95.1 | 96.5 |
| $\alpha = 0.15$ | ($\psi = 1$) | 95.6 | 94.9 | 95.2 | 95.3 | 95.4 | 95.3 | 95.2 | 95.0 |
| | ($\psi = 3$) | 95.2 | 94.9 | 94 | 96.8 | 95.2 | 95.2 | 94.9 | 96.0 |
| | ($\psi = 5$) | 95.5 | 94.4 | 95.6 | 96.8 | 96.0 | 95.2 | 95.6 | 96.3 |
| Asymmetric Trimming | | | | | | | | | |
| $q = 0$ | ($\psi = 1$) | 95.3 | 95.1 | 95.2 | 95.6 | 95.0 | 95.4 | 95.6 | 95.2 |
| | ($\psi = 3$) | 95.1 | 94.7 | 93.4 | 94.4 | 94.4 | 93.6 | 93.5 | 94.0 |
| | ($\psi = 5$) | 95.5 | 95.0 | 92.5 | 93.0 | 94.8 | 93.4 | 91.5 | 93.0 |
| $q = 0.01$ | ($\psi = 1$) | 94.7 | 94.5 | 94.8 | 94.3 | 94.7 | 94.5 | 95.2 | 94.8 |
| | ($\psi = 3$) | 92.0 | 94.3 | 92.8 | 91.8 | 92.7 | 94.2 | 92.8 | 92.0 |
| | ($\psi = 5$) | 94.3 | 94.7 | 93.9 | 94.8 | 94.0 | 94.9 | 93.8 | 94.6 |
| $q = 0.05$ | ($\psi = 1$) | 91.1 | 94.1 | 92.8 | 89.0 | 92.0 | 94.6 | 94.1 | 89.5 |
| | ($\psi = 3$) | 93.8 | 94.4 | 93.0 | 92.8 | 93.8 | 95.1 | 94.6 | 93.1 |
| | ($\psi = 5$) | 94.9 | 95.1 | 94.8 | 96.5 | 95.1 | 95.7 | 95.0 | 95.5 |

**Table 5**. Estimated differences (and 95% confidence intervals) in counterfactual probability in returning to work between minimally invasive myomectomy and minimally invasive hysterectomy in the first seven weeks[a].

| Week | IPTW (Estimator I) | | OW (Estimator I) | |
| --- | --- | --- | --- | --- |
| | Estimated Difference | Confidence Interval | Estimated Difference | Confidence Interval |
| 1 | 0.144 | (-0.127,0.415) | 0.073 | (0.015,0.131) |
| 2 | 0.274 | (0.039,0.509) | 0.211 | (0.128,0.295) |
| 3 | 0.346 | (0.157,0.534) | 0.254 | (0.161,0.347) |
| 4 | 0.319 | (0.144,0.494) | 0.262 | (0.170,0.354) |
| 5 | 0.237 | (0.069,0.405) | 0.187 | (0.095,0.279) |
| 6 | 0.198 | (0.040,0.356) | 0.176 | (0.086,0.265) |
| 7 | 0.095 | (-0.060,0.250) | 0.111 | (0.027,0.196) |

[a] The results are given by estimator I. Positive values indicate myomectomy leads to shorter time in returning to work.